\documentclass{webofc}
\usepackage[varg]{txfonts}   
\usepackage{xcolor}

\def\OO{\mathcal{O}}
\def\bb{{\boldsymbol b}}

\def\gs{g_{\mathrm{s}}}

\def\gsqb{g_{3\mathrm{d}}^2 b_\perp}

\def\gsq{g_{3\mathrm{d}}^2}
\def\gfour{g_{3\mathrm{d}}^4}
\def\gsix{g_{3\mathrm{d}}^6}

\def\mD{m_{\mathrm{D}}}
\def\mDsq{m_{\mathrm{D}}^2}

\def\bp{\bb_\perp}
\def\bbp{b_\perp}
\def\bps{b_\perp^2}

\def\qp{q_\perp}
\def\bqp{\boldsymbol q_\perp}

\def\Cbp{C(\bbp)}
\def\Cqp{C(\qp)}
\def\CQCDbp{C_\mathrm{QCD}(\bbp)}
\def\CEQCDbp{C_\mathrm{EQCD}(\bbp)}

\def\tension{\sigma_{\mathrm{EQCD}}}

\def\CR{C_\mathrm{R}}
\def\CA{C_\mathrm{A}}

\def\d{\mathrm{d}}

\def\qhn{\hat{q}_0}

\begin{document}

\title{ Non-perturbative phenomena in jet modification }

\author{\firstname{Guy D.} \lastname{Moore}\inst{1}\fnsep\thanks{\email{guy.moore@physik.tu-darmstadt.de}}\and
        \firstname{S\"oren} \lastname{Schlichting}\inst{2}\fnsep\thanks{\email{sschlichting@physik.uni-bielefeld.de}}\and
        \firstname{Niels} \lastname{Schlusser}\inst{1,3,4}\fnsep\thanks{\email{niels.schlusser@unibas.ch}} \and
        \firstname{Ismail} \lastname{Soudi}\inst{2,5}\fnsep\thanks{\email{isma@physik.uni-bielefeld.de}}
}

\institute{
Institut f\"ur Kernphysik, Technische Universit\"at Darmstadt\\
Schlossgartenstra{\ss}e 2, D-64289 Darmstadt, Germany
\and
Fakultät f\"ur Physik, Universit\"at Bielefeld\\ 
D-33615 Bielefeld, Germany
\and
Department of Physics \& Helsinki Institute of Physics\\
P.O. Box 64, FI-00014 University of Helsinki
\and
Biozentrum, Universität Basel\\
Spitalstrasse 41, 4056 Basel, Switzerland
\and
Department of Physics and Astronomy, Wayne State University\\
Detroit, MI 48201.
          }

\abstract{The interaction of a jet with the medium created in heavy-ion collisions is not yet fully understood from a QCD perspective. This is mainly due to the non-perturbative nature of this interaction which affects both transverse jet momentum broadening and jet quenching. 
We discuss how lattice simulations of Electrostatic QCD, can be matched to full, four dimensional QCD, to determine non-perturbative contributions to the momentum broadening kernel. We determine the momentum broadening kernel in impact parameter and momentum space and finally show how these results can be used in phenomenological calculations of in-medium splitting rates.
}

\maketitle

\section{Introduction}
One crucial signature of Quark-Gluon-Plasma (QGP) formation in ultra-relativistic heavy-ion collisions is the suppression of highly energetic particles or jets in the final state.
Before being detected, the hard partons must traverse the QGP and interact with the medium, leading to a substantial loss of their energy. In addition to the elastic interactions with the medium constituents, multiple soft scatterings between the hard partons and the medium trigger the partons to radiate. Since soft scatterings occur frequenctly, an infinite number of diagrams has to be resummed in order to compute effective $1\to2$ in-medium radiation rates. Different formalisms are employed in the literature to obtain the rate of radiation 
\cite{Gyulassy:1999zd,Gyulassy:2000er,Gyulassy:2003mc,Wiedemann:2000za,Salgado:2003gb,Arnold:2002ja,Djordjevic:2008iz,CaronHuot:2010bp}. While these formalisms differ on many aspects and approximations, they all rely on the description of the elastic scattering with the medium, obtained using the transverse momentum broadening kernel 
\begin{equation}    \label{C_q_perp}
    \Cqp \equiv \frac{(2\pi)^2 \d^3 \Gamma}{\d^2 \qp \, \d L} \, ,
\end{equation}
which describes the rate at which the hard partons exchange transverse momentum $\bqp$ with the medium. To obtain the broadening kernel, one can use various treatments of the medium, e.g.,~ if one treats the medium as many random, static, screened color centers, one obtains $\Cqp \propto \frac{1}{(\qp^2 + \mD^2)^2}$, with the Debye screening mass $\mD^2$. Furthermore, taking the medium to be made up of dynamical moving charges at leading order in perturbation theory \cite{Aurenche:2002pd} leads to $\Cqp \propto \frac{1}{\qp^2(\qp^2+\mD^2)}$. At very large parton energies, the interaction with the medium can be approximated as many individually small scatterings, leading to transverse momentum diffusion: $\Cbp \approx \hat{q} \, \bps/4$, also known as the harmonic oscillator approximation.
All these treatments rely on approximations and perturbative expansions, but the medium is quite strongly coupled and there can be large non-perturbative contributions even at high temperatures. It would be important to capture genuine non-perturbative contributions, in order to improve on the treatment of the jet-medium interactions.
For that sake, it has been found by Casalderrey-Solana and Teaney \cite{CasalderreySolana:2007qw} that the broadening kernel $C(\qp)$ can be related to its position-space ($\bbp$) version $\Cbp$ via the modified Fourier transform
\begin{equation}    \label{subtraction_FT}
\Cbp \equiv \int \! \frac{\d^2 \qp}{(2\pi)^2}
    \left( 1 - e^{i\bqp \! \cdot \, \bp} \right) C(\qp)  \, .
\end{equation}
$\Cbp$, in turn, can be computed from a light-like Wilson-loop operator with real-time techniques. Despite the jet being a high energy observable, its interactions with the medium are dominated by infrared contributions. The requirement of infrared resummation on the one hand and real-time techniques on the other hand pose a serious challenge. This issue can be resolved by treating the correlators to be slightly spacelike, making the infrared part of the calculation amenable to effective field theory (EFT) treatment within the framework of electrostatic QCD (EQCD) \cite{CaronHuot:2008ni}. To further support the convergence of the EQCD result, an evaluation on the lattice was pioneered \cite{Panero:2013pla} and finally extrapolated to the continuum \cite{DOnofrio:2014mld,Moore:2019lua,Moore:2019lgw}.

\section{Non-perturbative broadening kernel}

\subsection{Broadening kernel in position space}
\begin{figure}
    \includegraphics[width=\textwidth]{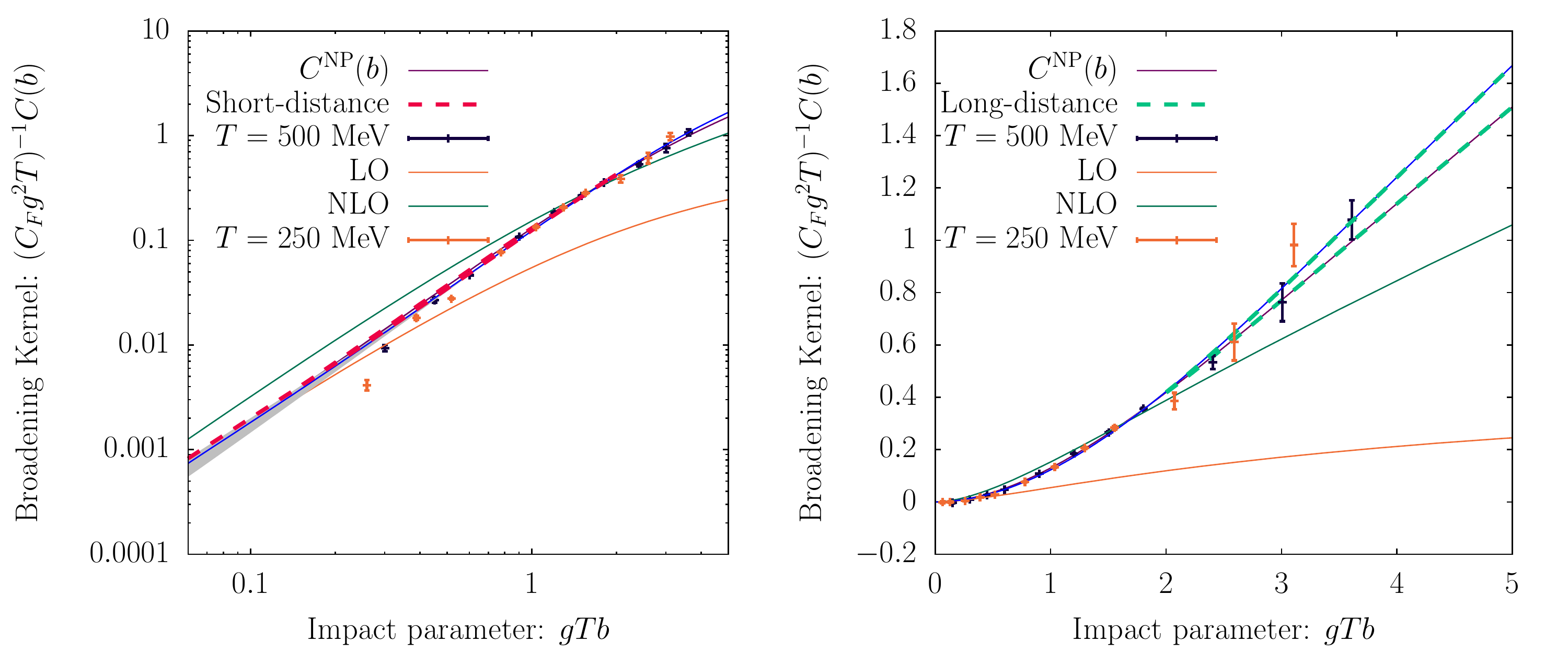}
    \caption{Non-perturbative elastic broadening kernel interpolation spline in the short-distance (left) and large-distance (right) regimes. We compare to both the short-distance limit from Eq.~(\ref{UV_limit}) and the long-distance limit from Eq.~(\ref{IR_limit}).}
    \label{fig:Spline}
\end{figure}
EQCD is a long-distance effective treatment of thermal QCD and therefore only reproduces the infrared (IR) regime of QCD well. In order to match the predictions of full QCD in the ultraviolet (UV) regime of $\Cqp$ (see \cite{Arnold:2008vd}), the incorrect EQCD UV limit has to be removed by subtraction and the correct UV limit from full thermal QCD has to be reinstated, such that following the strategy of \cite{Ghiglieri:2018ltw}
\begin{align}    \label{match_strategy1}
    \CQCDbp &= \left( \CQCDbp - \CEQCDbp \right) + \CEQCDbp \notag \\
            &= \left( C_\mathrm{QCD}^\mathrm{\OO(g^4)}(\bbp) 
                    - C_\mathrm{EQCD}^\mathrm{\OO(g^4)}(\bbp) \right)
                + C_\mathrm{EQCD}^\mathrm{latt}(\bbp) + \OO(g^6) \, ,
\end{align}
with the strong coupling constant $\gs$.
The difference between EQCD and full QCD in brackets is UV dominated and therefore safe to be evaluated perturbatively. The perturbative results are available at $\OO(g^4)$ in full QCD \cite{Arnold:2008vd} and EQCD \cite{CaronHuot:2008ni,Ghiglieri:2018ltw}. Continuum-extrapolated lattice data was computed in \cite{Moore:2019lgw}.
Since the lattice data is only available at finite $\bbp$, we have to rely on analytical information beyond that window of availability.

In the \textsl{limit of large impact parameter $\bbp$} -- small $\qp$ in transverse momentum space -- the dominant behavior should be linear in $\bbp$ caused by the Wilson loop following an area law \cite{Laine:2012ht}. These contributions are subject to a fit to the large-$\bbp$ tail of our lattice EQCD data, yielding the fit constant $A$ and the string tension $\tension$ from the literature \cite{Laine:2005ai}. The perturbative terms in \eqref{match_strategy1} can only contribute sub-dominantly, in fact, they give rise to a logarithmic term. Thus, the overall large-$\bbp$ limit reads
\begin{equation}	\label{IR_limit}
\frac{C_\mathrm{QCD}}{\gsq} (\bbp) \xrightarrow{\bbp \gg \; 1/\gsq} A + \frac{\sigma_\mathrm{EQCD}}{\gfour} \gsqb +  \frac{\gs^4 \CR}{\pi} \left[ \frac{y}{4} \left( \frac{1}{6} - \frac{1}{\pi^2} \right) + \frac{\CA}{8 \pi^2 \gs^2} \right] \log \Big( \gsqb \Big) \, ,
\end{equation}
in units of the three-dimensional coupling $\gsq \approx g^2 T$, with the dimensionless  screening mass ratio $y \equiv \frac{\mDsq}{\gfour} \Big\vert_{\mu = \gsq}$ and the Casimir operators $\CR$ of the representation $R$ of the jet particle and $\CA$ the Casimir of the adjoint representation.

The \textsl{opposite limit} is governed by the full QCD perturbative part of \eqref{match_strategy1}, since lattice EQCD and EQCD perturbation theory agree in the limit of small impact parameter. Dominated by the harmonic oscillator term $\Cbp \approx \hat{q} \, \bbp^2/4$, subleading effects due to screening occur at $\OO(\bbp^2 \ln \bbp)$. Altogether, the result reads
\begin{equation}	\label{UV_limit}
\frac{C_\mathrm{QCD}}{\gsq} (\bbp) \xrightarrow{\bbp \ll \; 1/\mD} - \frac{\CR}{8 \pi} \frac{\zeta(3)}{\zeta(2)} \left( -\frac{1}{2 \gs^2} + \frac{3 y}{2} \right) \gfour \bbp^2 \log \Big( \gsqb \Big) + \frac{1}{4} \frac{\qhn}{\gsix} \gfour \bbp^2 \, ,
\end{equation}
with the scale-independent part of the jet-quenching parameter $\qhn$.

We present in Fig.~\ref{fig:Spline} the fully matched results for $T=250$MeV and $500$MeV in orange and black points, respectively. Using the data points obtained and the limiting behaviors of Eqns.~(\ref{IR_limit}-\ref{UV_limit}), we construct a continuous spline for each temperature as discussed in \cite{Moore:2021jwe} shown as a violet line for $T=250$MeV and as a blue line for $T=500$MeV.
Overall, we observe that even though the non-perturbative results are obtained at different temperatures and coupling constants, they scale onto very similar broadening kernels when we consider the broadening kernel in units of $g^2T$ as a function of impact-parameter $\bbp$ in units of $\frac{1}{gT}$.
The right panel of Fig.~\ref{fig:Spline} shows how the long-distance limit is approached by the data points and splines obtained from fully matched lattice results at two different temperatures. Moreover, purely perturbative estimates of $\Cbp$ at leading (LO) and next-to-leading order (NLO) are shown as a benchmark. 
The left panel of Fig.~\ref{fig:Spline} displays how the interpolating curve $C^\mathrm{NP}(\bbp)$ approaches the correct small-$\bbp$ limit from \eqref{UV_limit} and still matches the fully matched data points based on lattice EQCD as well as possible. The perturbative results approach the same short-distance behavior at different scales.

\subsection{Broadening kernel in momentum space}
\begin{figure}
    \centering
    \includegraphics[width=0.6\textwidth]{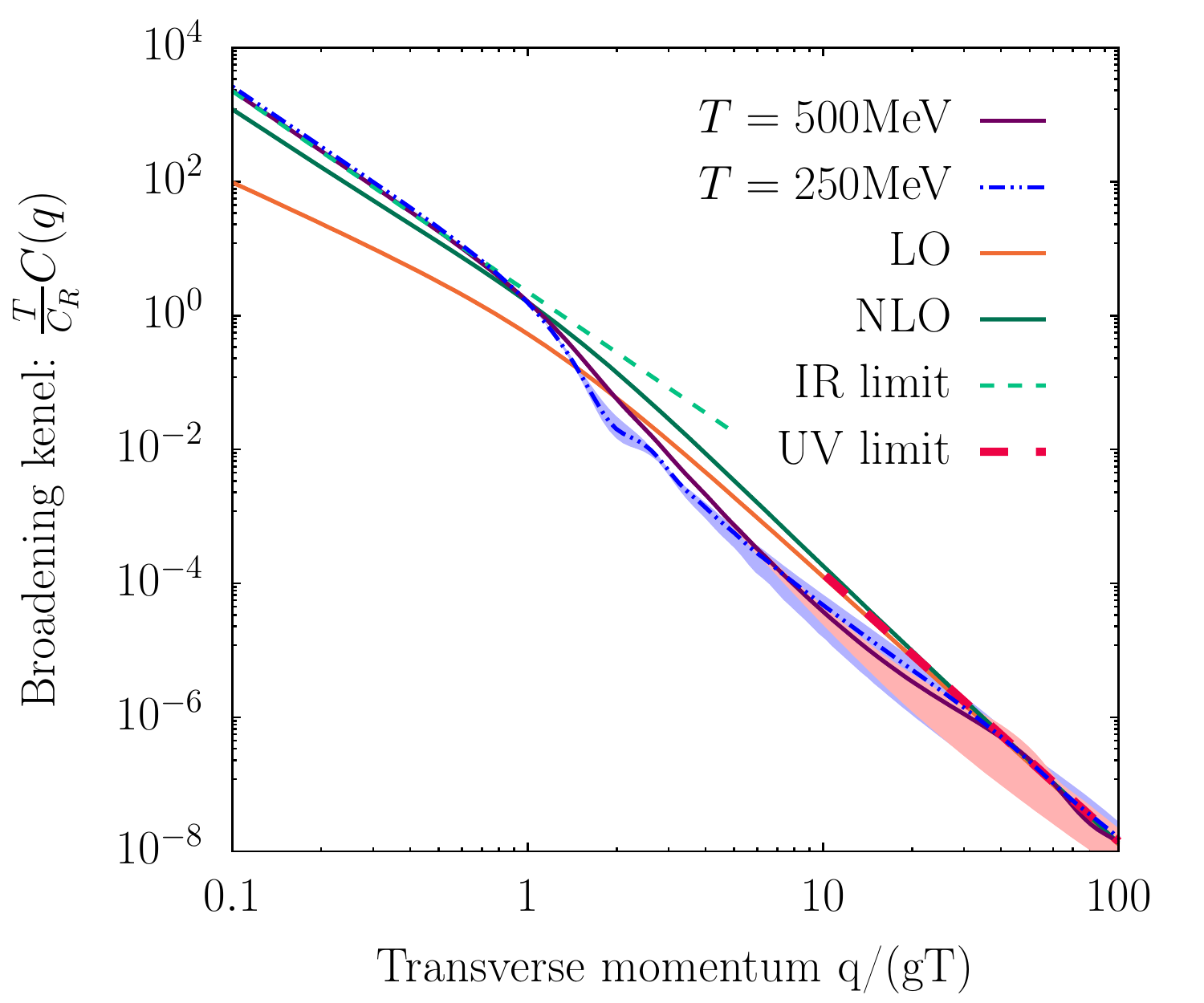}
    \caption{Elastic broadening kernel in momentum space for both $T=250,500$MeV, where the blue and red bands represent the error in the choice of the spline for $250$MeV and $500$MeV respectively. We compare to both the UV and IR limit, given as the Fourier transforms of Eqs.~\ref{UV_limit} and \ref{IR_limit}, see \cite{Schlichting:2021pwx}. }
    \label{fig:InverseKernel}
\end{figure}
The lattice EQCD calculation is performed in impact-parameter space ($\bp$), which is of direct use for obtaining splitting rates in a medium of infinite size. However, in order to obtain splitting rates in a realistic medium of finite size, it is more convenient to work in momentum space ($\qp$). Therefore, we proceed to Fourier transform the resulting fully matched $\Cbp$ to momentum space. 
The inverse Fourier transform seems to be straightforward to compute, however, since the data points are sparse and the kernel is divergent at large impact parameter, performing the highly oscillatory integrals involved is rather challenging. We can evade these difficulties, if we instead Fourier transform the coordinate space derivative $\frac{\d C(\bp)}{\d b_\perp}$ of the momentum broadening kernel. Using Eq.~(\ref{subtraction_FT}), one can write
\begin{align}    
    C(\qp)  =& \frac{2\pi}{\qp}\int_0^{\infty} \! \d b_\perp~  b_\perp J_1(b_\perp\,\qp) \frac{\d C(b_\perp)}{\d b_\perp}\, .
\end{align}
Numerical details of the integration are discussed in \cite{Schlichting:2021pwx} and we present the resulting broadening kernel in momentum space ($\qp$) in Fig.~\ref{fig:InverseKernel} compared to perturbative broadening kernels. We observe how both the non-perturbative and NLO results display similar behavior $(\propto1/\qp^3)$ in the IR limit which is the Fourier transformation of the linear long-distance behavior in Eq.~(\ref{UV_limit}), however, they come with different prefactors.
Conversely, all the different kernels recover the UV behavior $(\propto1/\qp^4)$ at large momentum $\qp$, which is associated with hard scatterings.

\section{Splitting rates in an infinite medium}
\begin{figure}[h!]
    \includegraphics[width=0.98\textwidth,trim={ 0 14.1cm 0 14.3cm},clip]{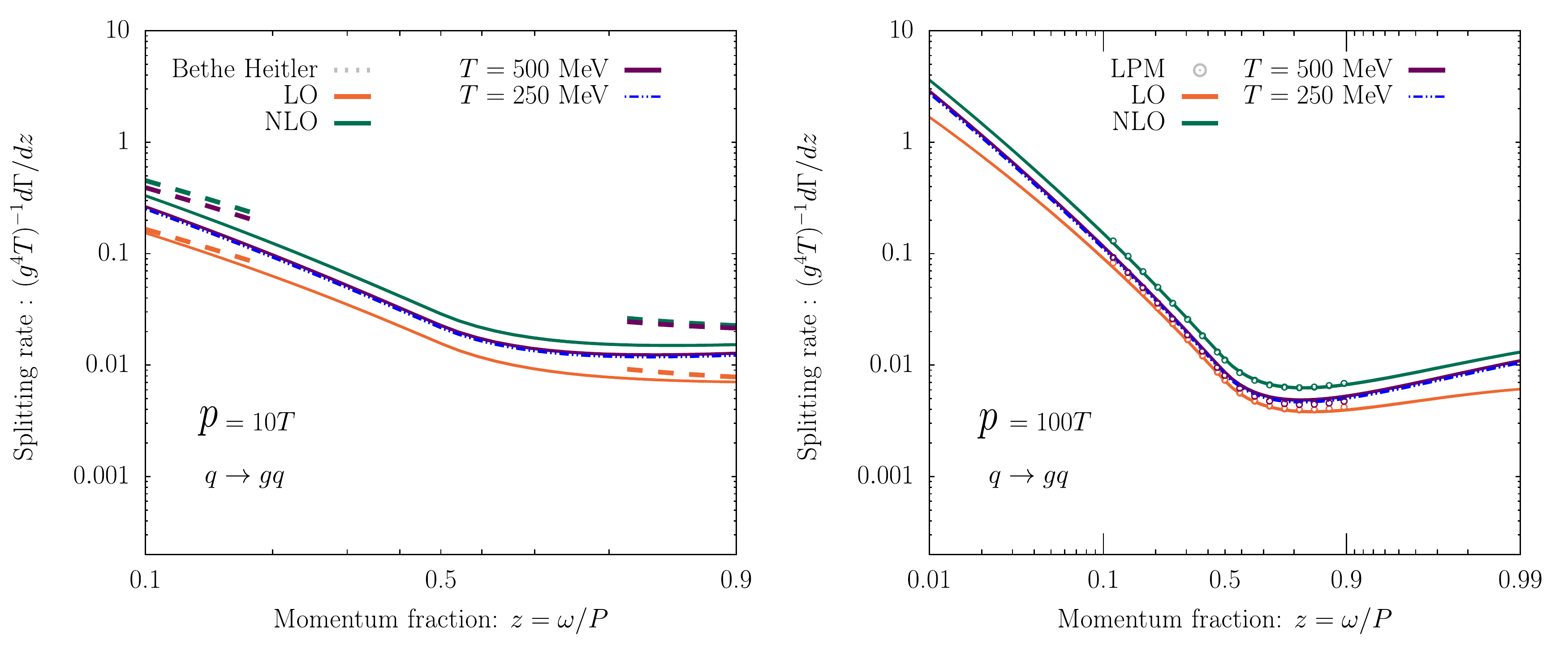}
    \caption{In-medium splitting rate of gluon by a parent quark in a medium of temperature $T=250$MeV (dashed blue lines) and $T=500$MeV (full purple lines). Different columns correspond to parent energies $P=10T$ (left) and $P=100T$ (right). We compare with rates computed using the perturbative leading order (orange) and next-to-leading order (green) elastic broadening kernels. The Bethe-Heitler rates and LPM rates are shown with dashed lines and circles, respectively, using the color of the corresponding kernel. }
    \label{fig:SplittingRates500}
\end{figure}

While there are different formalisms used to compute medium induced radiation in QCD plasma \cite{Baier:1996kr,Zakharov:1996fv,Gyulassy:2000er,Arnold:2001ms}, here we follow the formalism of Arnold, Moore and Yaffe (AMY) \cite{Arnold:2001ms}, which can directly make use of the broadening kernel in impact-parameter space ($\bp$) and provides an effective rate $\d \Gamma_{ij}/ \d z(P,z)$, which corresponds to the rate at which particle $i$ with energy $P$ radiates particle $j$ with energy $\omega = zP$ in an infinite medium. We refer the reader to \cite{Moore:2021jwe} for details of the formalism and the numerical procedure. 

Our results are presented in Fig.~\ref{fig:SplittingRates500}, where in addition to the splitting rates obtained using the non-perturbative broadening kernels at $T=250,500$MeV, we compare to the rates obtained by employing leading and next-to-leading order results for momentum broadening \cite{Moore:2021jwe}. Furthermore, we show the Bethe-Heitler rates (dashed lines) and the deep LPM rates (circles) which are semi-analytical approximations of the rate calculation at small typical momentum $Pz(1-z)\ll T$ and large typical momentum $Pz(1-z)\gg T$, respectively (c.f.~ for details of the derivation \cite{Moore:2021jwe}). We find that the two temperatures considered for the non-perturbative results do not display remarkable difference when the rate is measured in units of $[g^4T]$, which is a result of the scaling behavior we observed in the broadening kernel. The LPM suppression is recovered at large typical energies, where the splitting rates match the deep LPM rates, while the unsuppressed Bethe-Heitler rates are only reached at low typical momentum beyond the range of momentum fraction shown. When comparing to the rates obtained using perturbative broadening kernels, we observe that at high momentum $Pz(1-z)\gg T$ the rates are closer to the LO results, since this region probes the short-distant behavior of the broadening kernel where the non-perturbative kernel is matched to the LO one. Conversely, at low momentum $z(1{-}z)E \ll T$, where long-distance region is more relevant, the non-perturbative rate is closer to the NLO rate since they feature similar linear behavior at large impact parameter.

\section{Splitting rates in a finite size medium }
\begin{figure}
    \centering
    \includegraphics[width=0.6\textwidth]{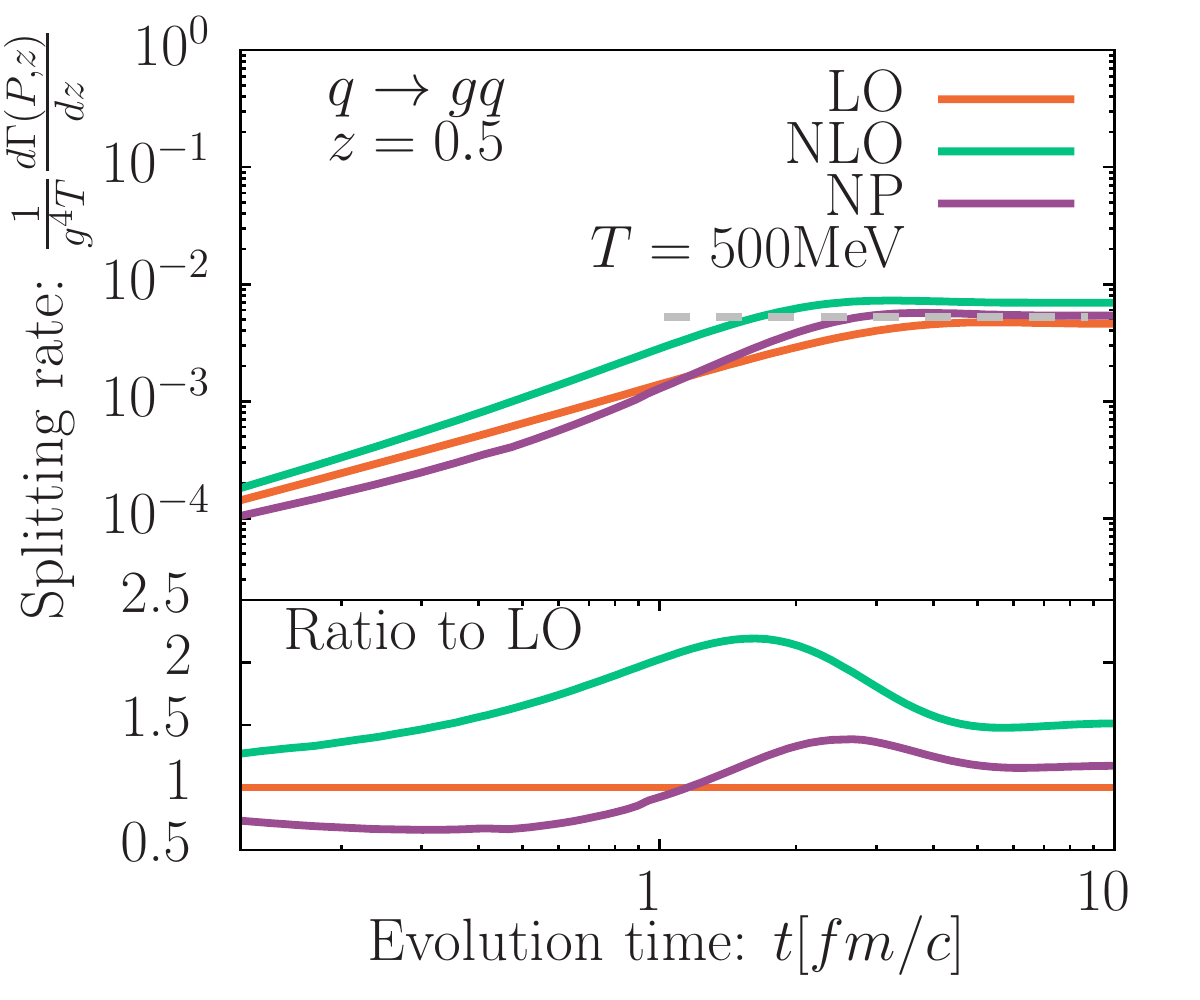}
    \caption{Medium-induced splitting rate of a gluon with momentum fraction $z=0.5$ from a parent quark with energy $P=300T$ in an equilibrium medium with temperature $T=500$MeV as a function of the evolution time $t$. We compare calculation done using the different collisional broadening kernel as shown in Fig.~\ref{fig:InverseKernel} (the temperature and coupling constant for the perturbative results are matched to the $T=500$MeV data in Tab.~1 of \cite{Moore:2021jwe}). The lower panel displays the ratio to the LO results. }
    \label{fig:FiniteMediumVSKernels}
\end{figure}
Using the broadening kernel $C(\qp)$ in momentum space, following \cite{CaronHuot:2010bp}, we are able to obtain the splitting rates of medium-induced radiation in a plasma of finite size. 
In \cite{Schlichting:2021pwx}, we provide details of the procedure we employ to obtain the splitting rate $\d \Gamma_{ij}/ \d z(P,z,t)$, which describes the rate at which particle $i$ with energy $P$ radiates particle $j$ with energy $\omega=zP$ after a time $t$ inside the medium. In Fig.~\ref{fig:FiniteMediumVSKernels}, we display results of quark with energy $P=300T$ radiating a gluon with momentum fraction $z=0.5$ as a function of the evolution time $t$ the quark spends in the medium. We present results at $T=500$MeV using the obtained non-perturbative broadening kernel (NP) as well as the Leading Order (LO) and Next-to-Leading Order (NLO) perturbative broadening kernels. Dashed gray line represent the infinite medium results obtained in the previous section. We also show in the lower panel the ratio with respect to the LO results. At early times, the splitting rates display a linear behavior and quickly saturate at late times when the rate recovers the infinite medium results. 
We observe that the non-perturbative rate in fact starts lower than the LO before it evolves in between the LO and NLO.

\section{Discussion}
In this work, we considered non-perturbative contributions to the momentum broadening kernel $C(\qp)$ calculated in lattice EQCD \cite{Moore:2019lgw}. Since EQCD is a long-distance effective theory of QCD, we supplied our lattice data with the correct short-distance behavior to obtain the non-perturbative momentum broadening kernel in QCD valid at all scales. Employing this kernel in impact parameter space, we computed medium-induced splitting rates by a hard parton traversing a QCD plasma of infinite size. We compared the resulting splitting rates to ones obtained using leading and next-to-leading order perturbative broadening kernels. We find that while the LO results are more relevant for hard momentum ($Pz(1-z)\gg T$) where the short-distance limit is more relevant, the rates display a similar behavior to the NLO for soft momentum ($Pz(1-z)\gg T$).
Although the perturbative series for the broadening kernel $\Cbp$ is apparently not convergent \cite{CaronHuot:2008ni}, we observe that the splitting rates for the non-pertubatively determined $\Cbp$ mostly fall between the LO and NLO results. 

Beyond the splitting rates in a medium of infinite size, we were able to transform the non-perturbative momentum broadening kernel $C(\qp)$ to momentum space, which is more favorable to compute splitting rates in media of finite extent. Following \cite{CaronHuot:2010bp}, we computed splitting rates in a finite size QCD medium, and compared our results to calculations with the leading order and next-to-leading order perturbative broadening kernels. We observe that while the rates from NLO display a large difference from the LO rates, non-perturbative results do not deviate beyond a band of $50\%$ around the LO. 

Since medium-induced emissions are a dominant mechanism of jet energy loss in QCD plasmas, it will be important to incorporate our results in phenomenological studies of jet quenching. Furthermore, there has been ongoing progress in obtaining non-perturbative contributions to the thermal masses \cite{Moore:2020wvy} and we envision including these results in future work.

\bibliography{biblio}

\end{document}